\documentclass[]{jfm}

\usepackage{graphicx}
\usepackage{xcolor}
\usepackage{bm}
\usepackage{epstopdf,epsfig}
\usepackage{newtxtext}
\usepackage{newtxmath}
\usepackage{natbib}
\usepackage{hyperref}
\usepackage{comment}
\usepackage{amsmath}
\usepackage{mathtools}

\hypersetup{
    colorlinks = true,
    urlcolor   = blue,
    citecolor  = black,
}

\newcommand{\RomanNumeralCaps}[1]

\newcommand{\vf}{\bm{f}}

\newcommand{\vphi}{\bm{\phi}}

\newcommand{\vQ}{\bm{Q}}
\newcommand{\vF}{\bm{F}}
\newcommand{\vG}{\bm{G}}

\title{Surface instabilities in laminar compressible boundary layers with sublimation}

\author{Blaine Vollmer\aff{1}
  \corresp{\email{blainev2@illinois.edu}},
  Alberto Padovan,\aff{1}
 \and Daniel J. Bodony\aff{1}}

\affiliation{\aff{1}Department of Aerospace Engineering, University of Illinois Urbana-Champaign, Urbana, IL 61801, USA}

\begin{document}
\maketitle

\begin{abstract}
Surface patterns on ablating materials are known to appear in both high-speed ground and flight tests, but the mechanisms behind their formation are not known.
In this paper, the origins of surface patterns are investigated via a local linear stability analysis of compressible laminar boundary layers over a flat camphor plate.
The effects of sublimation and conjugate heat transfer are included both on the baseflow and the linear fluctuations.
This newly developed framework identifies a single mode that fully characterizes the stability of the surface, and this surface mode becomes unstable under laminar conditions only when the wall temperature exceeds that of an adiabatic wall, $T_{ad}$.
These findings are consistent with experimental observations, where laminar flow conditions at adiabatic wall temperatures are observed to be stable.
The present analysis also reveals that the nature of this surface mode varies as a function of the oblique angle $\psi = \tan^{-1}{\beta/\alpha}$, where $\alpha$ and $\beta$ are the streamwise and spanwise wavenumbers.
As the wall temperature increases, the most unstable orientation of the surface mode shifts from streamwise alignment ($\psi = 0$), toward the sonic angle ($\psi = \psi_s = \cos^{-1}(1/M_e)$), and then towards spanwise alignment ($\psi = 90^\circ$).
Finally, a critical wavenumber is identified (i.e., one at which the temporal growth rate reaches a maximum) which implies the formation of a surface pattern of a specific wavelength and orientation.

\end{abstract}


\section{Introduction} \label{sec:introduction}
\begin{figure}
    \centering
    \includegraphics[width=0.2\textwidth]{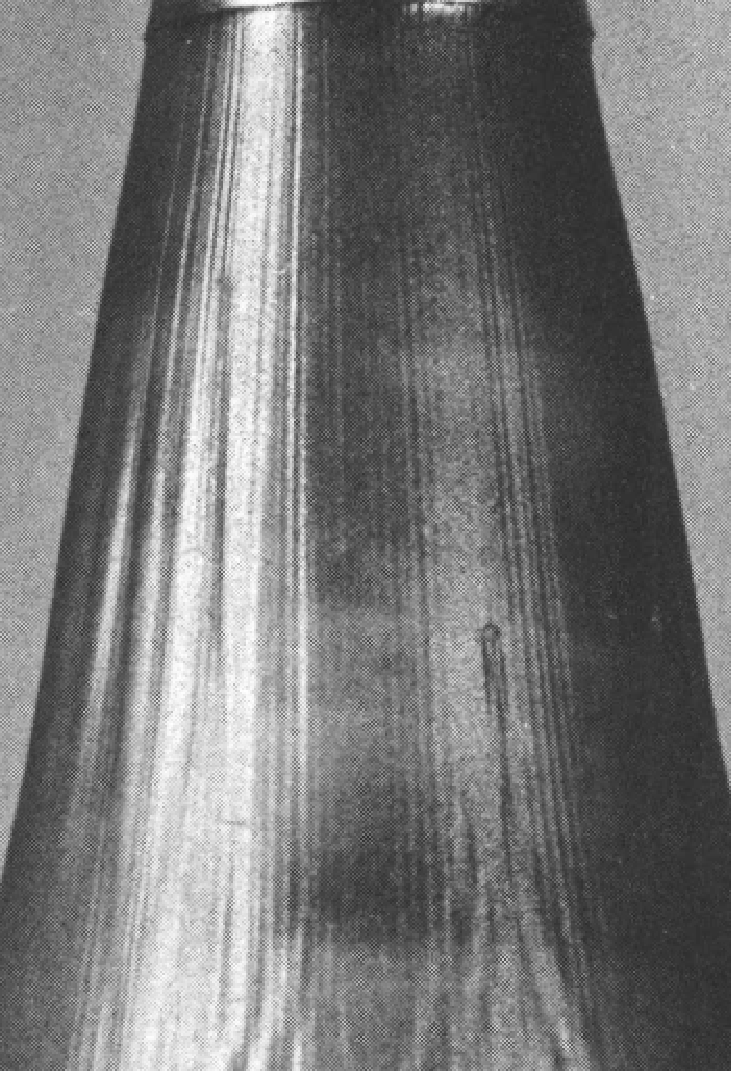}
    \includegraphics[width=0.2\textwidth]{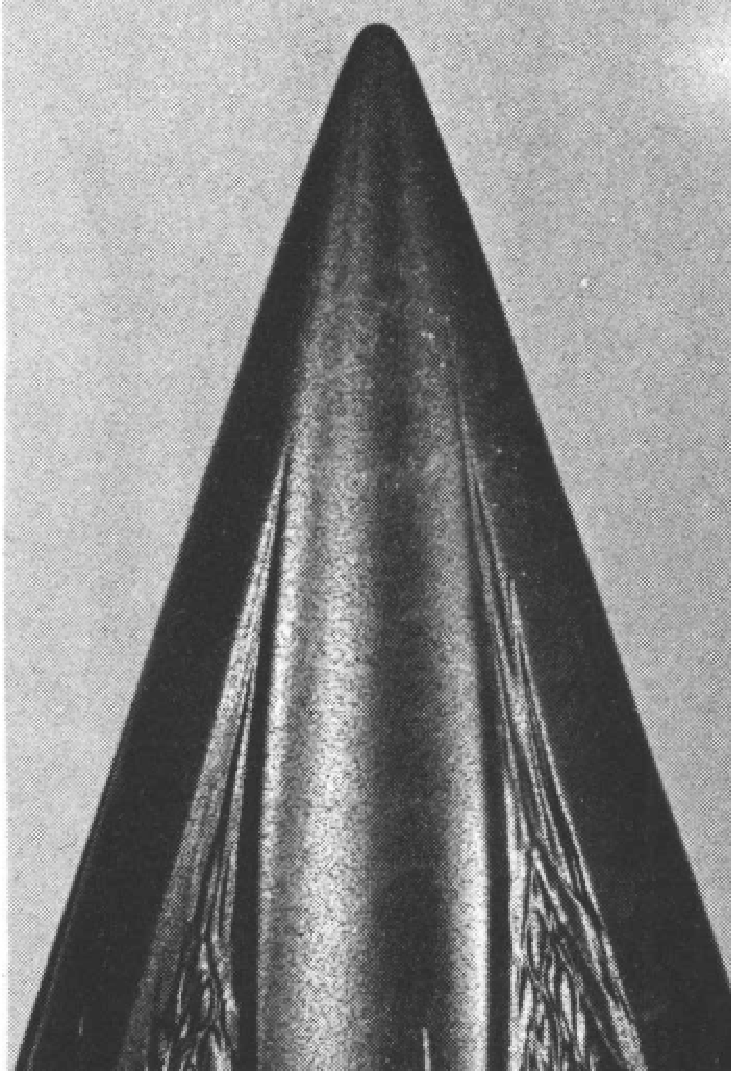}
    \includegraphics[width=0.2\textwidth]{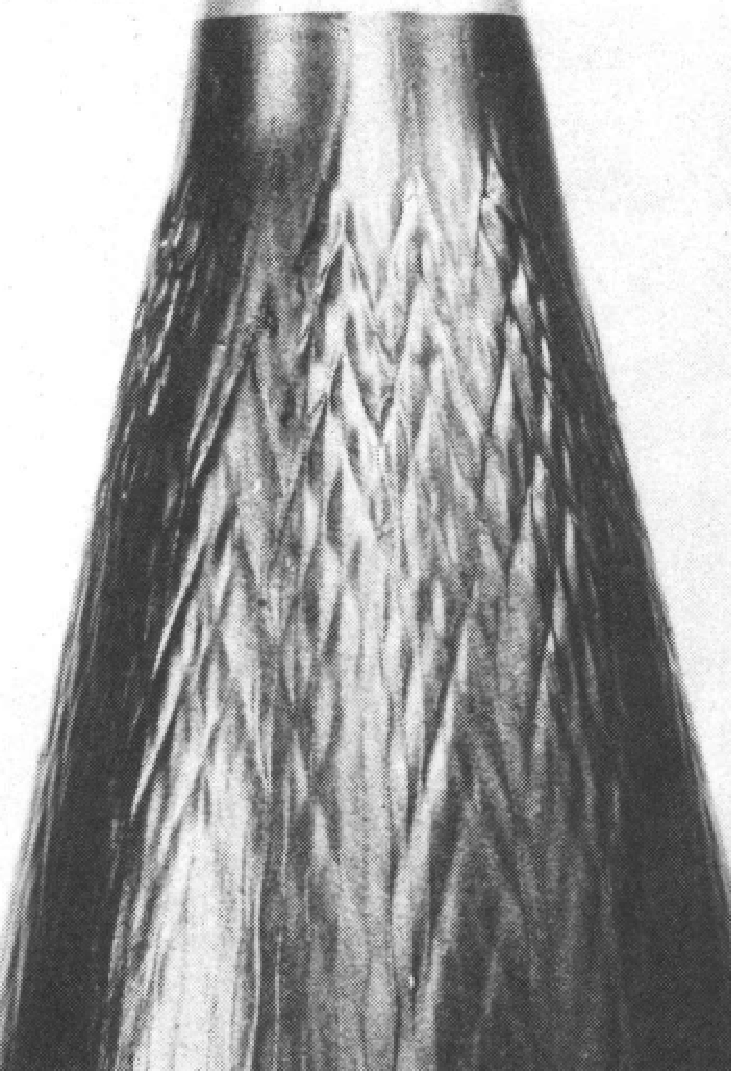}
    \caption{Ablation patterns showing streamwise grooves (left), turbulent wedges (middle), and crosshatching (right). Images taken from Figure 1 of \citet{stock1973cross}.}
    \label{fig:surface pattern images}
\end{figure}

Ablating thermal protection systems are used to shield hypersonic vehicles from the thermal loads experienced during flight. 
As the surface ablates, the vehicle surface changes shape and can develop surface patterns.
Both flight and ground testing have shown that ablating materials can exhibit macro-scale surface patterns of different types, including streamwise grooves, wedges, crosshatching, and scallops \citep{grabow1975surface}. 
These patterns can occur for melting, sublimating, and charring processes, and are observed on a wide range of materials including phenolics, plastics, and wood \citep{stock1973hypersonic}.
Examples of sublimation-induced surface patterns are shown in Figure \ref{fig:surface pattern images}.
The crosshatched surface (right panel in Figure \ref{fig:surface pattern images}) has received the most attention in the literature
and it has been shown that the crosshatch pattern angles correlate well with the local Mach angle \citep{swigart1974cross}.

Experiments have shown that surface patterns can develop on ablating materials when the flow is both supersonic and turbulent.
Such patterns have not been observed under laminar supersonic or turbulent subsonic conditions.
In addition, all tests have been performed under adiabatic ($T_w=T_{ad}$) or cold-wall ($T_w<T_{ad}$) conditions as a result of high stagnation temperatures and long exposure times \citep{stock1973cross}.
A summary of these experimental results is provided in Table \ref{tab:summary of experiments}.
\begin{table}
  \begin{center}
\def~{\hphantom{0}}
  \begin{tabular}{c@{\hskip 20pt}c@{\hskip 20pt}c}
         & \raisebox{1ex}{Cold Wall ($T_w \leq T_{ad}$)}  &   \raisebox{1ex}{Hot Wall ($T_w>T_{ad}$)} \\
       Laminar   & Unconditionally Stable & Untested \\
       Turbulent   & Conditionally Unstable & Untested \\
  \end{tabular}
  \caption{Summary of experimental surface pattern observations.}
  \label{tab:summary of experiments}
  \end{center}
\end{table}
\\
While surface patterns have only been observed under turbulent flow conditions, an exploration of surface stability under laminar conditions serves as a useful foundation for future study and may reveal new surface patterns and mechanisms.
We reiterate that while the observed patterns shown in Figure \ref{fig:surface pattern images} provide the initial motivation for studying linear surface instabilities, our study of surface stability goes beyond these observations to investigate general linear periodic surface patterns.
In addition to laminar flows, the study of hot wall conditions is warranted not only to fully characterize the behavior of ablating surfaces and improve the understanding of surface stability mechanisms, but also because hot walls may occur during reentry.
For example, as a vehicle decelerates, the stagnation temperature decreases with Mach number, which can create a condition in which the surface temperature remains greater than the surrounding flow.
If the conditions for surface pattern formation are well understood, they can be used to design vehicles and trajectories that prevent their formation.

\begin{figure}
    \centering
    \includegraphics[trim=30 1 1 1, clip, width=0.65\textwidth]{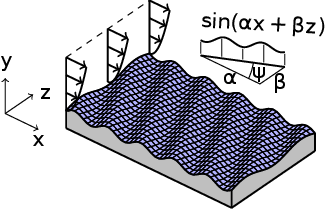}
    \caption{Schematic of local linear stability of an ablating surface}
    \label{fig:linear stability diagram}
\end{figure}

The mechanism by which surface patterns emerge on ablating materials remains unknown. 
\cite{swigart1974cross} reviewed several proposed explanations for crosshatching, including differential ablation, inelastic deformation, and liquid layer models. 
While some of these mechanisms could account for some of the observed patterns, the authors concluded that further investigation is needed.
The present work focuses on the differential ablation of pure sublimators using a local linear stability approach. 
A schematic of the system is shown in Figure \ref{fig:linear stability diagram}.
The framework and methodology presented here may be readily extended to porous ablators and inelastic deformation models to study alternative mechanisms.

The differential ablation mechanism was studied by \cite{lees1972stability}, where a simplified linear stability analysis of the coupled fluid-sublimation system was performed on a mean turbulent baseflow.
The simplifying assumptions include steady fluid perturbations, isothermal boundary conditions, single-species chemistry, and non-sublimating baseflows. 
The use of an isothermal wall boundary condition means that sublimation is not appropriately accounted for, and the sublimation rate was instead found by the energy balance that produces an isothermal wall.
The linear analysis by \cite{lees1972stability} identified an oblique surface mode as the most unstable, demonstrating that differential ablation can produce surface instabilities. 
However, \cite{swigart1974cross} reported another result from Lane and Ruger (unpublished), that contradicted the results of \cite{lees1972stability}, claiming to show no such instability. 
The Lane and Ruger results do not appear to be published, and it is still unknown whether differential ablation is a viable mechanism for surface pattern formation.

This paper presents a more complete formulation of differential ablation stability than \cite{lees1972stability} by considering the full unsteady system with sublimation models and conjugate heat transfer within the solid.
The surface stability of supersonic laminar flow over a flat camphor surface is analyzed at varying wall temperature ratios.
The surface is found to be unstable for wall temperatures sufficiently greater than the adiabatic temperature, $T_w > 1.05~ T_{ad}$, and stable otherwise.
These findings are consistent with experimental observations where patterns do not develop for laminar cold-wall conditions, and the analysis predicts that hot wall experiments--which to our knowledge have not yet been performed--will produce a laminar instability.
The governing equations for the fluid, solid, and interface sublimation are presented in Section \ref{chapter:governing eqns}.
Section \ref{chapter:linearization} presents the linearization procedure, and Section \ref{chapter:Results} contains the linear stability results.

\section{Governing Equations and Material Models} \label{chapter:governing eqns}
\subsection{Fluid}
The fluid is governed by the frozen-flow multi-species compressible Navier-Stokes equations, written in compact form as
\begin{equation}
    \frac{\partial \vQ}{\partial t} + \frac{\partial \vF_i}{\partial x_i} + \frac{\partial \vG_i}{\partial x_i} = 0,
  \label{Navier-Stokes}
\end{equation}
where $t$ is time, $x_i$ are the spatial coordinates, $\vQ$ is the vector of conservative flow variables,~$\vF$ denotes the inviscid fluxes and $\vG$ denotes the viscous fluxes,

\begin{equation}
\setlength{\arraycolsep}{0pt}
\renewcommand{\arraystretch}{1.3}
\vQ = \left[
\begin{array}{ccccc}
  \rho    \\
  \rho u_1 \\
  \rho u_2 \\
  \rho u_3 \\
  \rho E \\
  \rho_1 \\ 
  \vdots \\
  \rho_{n-1}
\end{array}  \right] , \quad
\vF_i = \left[ 
\begin{array}{ccccc}
  \rho u_i    \\
  \rho u_1 u_i + p \delta_{1i} \\
  \rho u_2 u_i + p \delta_{2i} \\
  \rho u_3 u_i + p \delta_{3i} \\
  \left( \rho E + p \right) u_i \\
  \rho_1 u_i \\ 
  \vdots \\
  \rho_{n-1} u_i
\end{array}  \right] , \quad
\vG_i = \left[ 
\begin{array}{ccccc}
  0    \\
  -\tau_{1i} \\
  -\tau_{2i} \\
  -\tau_{3i} \\
  -u_j \tau_{ij} + q_i \\
  \rho_1 w_{1i} \\
  \vdots \\
  \rho_{n-1} w_{(n-1)i}
\end{array}  \right].
\label{NS_vectors}
\end{equation}
Here, $\rho$ is density, $u_i$ are the Cartesian velocity components, $E$ is the total energy, $p$ is the pressure, and $\rho_k$ and $w_{ki}$ are mass-density and $i$th velocity for each chemical species $k \in\{1,2,\ldots,n\}$. 
The viscous stress tensor $\tau_{ij}$, heat-flux vector $q_i$, and species velocities are defined as
\begin{equation}
    \tau_{ij} = \mu \left(\frac{\partial u_i}{\partial x_j} + \frac{\partial u_j}{\partial x_i} \right) + \lambda \frac{\partial u_i}{\partial x_i} \delta_{ij}, \quad
    q_i = -\kappa \frac{\partial T}{\partial x_i} + \sum_{k=1}^n \rho_k h_k w_{ki}, \quad
    w_{ki} = -\frac{D_k}{c_k} \frac{\partial c_k}{\partial x_i}
  \label{Transport}
\end{equation}
where $T$ is the fluid temperature, $h_k$ is the species enthalpy, and $c_k = \rho_k /\rho$ are the species mass fractions. 
The total energy per unit volume is given by
\begin{equation}
    \rho E = \frac{p}{\gamma-1} + \frac{1}{2}\rho \sum_{i=1}^3 u_i^2 + \sum_{k=1}^n \rho_k h_k^{0}
  \label{Total Energy}
\end{equation}
where $h_k^0$ is the enthalpy of formation. 
The ideal gas law is used for the mixture,
 \begin{equation}
     p = \sum_{k=1}^n \rho_k R T ,
   \label{Ideal gas}
 \end{equation}
where $R$ is the mixture gas constant.

The transport coefficients ($\mu_k$, $\kappa_k$, and $D_k$) for a single species are given by kinetic theory \citep{hirschfelder1964molecular}. 
The mixture transport coefficients are found using the Wilke mixing rule with Eucken correction for conductivity. 
Specific model details are provided in Appendix \ref{appendix A}.

\subsection{Solid}
For the solid phase, a non-porous sublimator comprised of a single material with spatially-homogeneous isotropic thermal conductivity is considered.
The solid is governed by the heat equation
\begin{equation}
   \frac{\partial T_s}{\partial t} = \alpha_s \nabla^2 T_s ,
\label{Conduction}
\end{equation}
where $T_s$ is the temperature of the solid and $\alpha_s$ the thermal diffusivity. 

\subsection{Interface}
The interface between the fluid and sublimator is governed by mass and energy balances across the surface. 
These interface conditions can be derived formally from the governing equations on either side of the interface as done by \cite{padovan2024extended}.
For an $x_2$-normal interface that does not accumulate mass, the interface conditions for mass conservation reduce to
\begin{equation}
    \left(\rho_k v \right)_f + \left(\rho_k w_{k2} \right)_f = \dot{m}_k,
    \label{mass balance - species}
\end{equation}
where $v_f=u_{2,f}$ is the vertical velocity component at the wall and $\dot{m}_k$ is the species mass flux produced by the surface reaction. Subscript $f$ refers to the external fluid.
Summing over all species we find
\begin{equation}
    \dot{m} = \left(\rho v \right)_f.
\label{mass balance}
\end{equation}
Similarly, the surface energy balance gives:
\begin{equation}
    \left(-\kappa \frac{\partial T}{\partial y} \right)_f + \left(\rho v h \right)_f + \left( \sum_{k=1}^n \rho_k w_{k2} h_k \right)_f = \left(-\kappa \frac{\partial T}{\partial y} \right)_s + \dot{m}h_s
\label{energy balance}
\end{equation}
where $h_s$ is the enthalpy of the solid sublimator, which accounts for surface reactions. 
For a binary mixture, with $k=1$ as the sublimator, this can be simplified to:
\begin{equation}
    \left(-\kappa \frac{\partial T}{\partial y} \right)_f = \left(-\kappa \frac{\partial T}{\partial y} \right)_s - \dot{m}h_{sg}
\label{energy balance - simple}
\end{equation}
where $h_{sg}$ is the enthalpy of sublimation (solid to gas).
The rate of surface recession is found by the rate of mass lost to sublimation
\begin{equation}
    \dot{m} = -\rho_s v_s,
\label{surface velocity}
\end{equation}
where $\rho_s$ is the density of the solid and $v_s$ is the velocity of the surface.
The sublimation rate is found from the Knudsen-Langmuir model in combination with the Clausius-Clapeyron relation for vapor pressure using the triple point:
\begin{equation}
    \dot{m} = a_1 \frac{p_1^{sat} - p_1}{\sqrt{2 \pi R_1 T_w}}, \quad\ln \left(\frac{p_1^{sat}}{p_{TP,1}} \right) =  \frac{h_{sg}}{R_1}\left(\frac{1}{T_{TP,1}} - \frac{1}{T_w} \right)
\label{Knudsen-Langmuir}
\end{equation}
where $a_1$ is the accommodation coefficient and $p_1$ is the partial pressure of the sublimating material.
The superscript $sat$ refers to the saturation pressure, and the subscript $TP$ refers to the triple point.

\subsubsection{Receding Surface} \label{subsection:receding surface}
The governing equations will be applied to the uniformly receding baseflow coordinate system, while perturbations to the surface are handled by linearized boundary conditions. 
The governing equations are transformed from the stationary reference frame $x_i$ to a coordinate system translating with the receding surface, $\bar{x}_i$.
The time derivative in \eqref{Navier-Stokes} transforms to
\begin{equation}
    \frac{\partial \vQ}{\partial t}\bigg\rvert_{x_i} = \frac{\partial \vQ}{\partial t}\bigg\rvert_{\bar{x}_i} - v_s \frac{\partial \vQ}{\partial x_2},
  \label{Navier-Stokes Transformation}
\end{equation}
where $v_s$ is the surface velocity in the vertical direction, $x_2$.
Since $v_s$ is constant for the uniformly receding baseflow, the moving coordinate term is combined with the convective fluxes to convert the $u_2$ velocity to the relative velocity $u_2 - v_s$. 
To analyze this effect, Equations \eqref{mass balance} and \eqref{surface velocity} are combined to give
\begin{equation}
    v_f = -v_s \left( \rho_s / \rho_f + 1 \right)
\end{equation}
where $\rho_f$ and $v_f$ are the fluid density and velocity at the wall. 
Sublimators typically have a density much larger than the fluid such that $\rho_s \gg \rho_f$, which gives $v_f \gg v_s$, and the relative velocity terms can be neglected. 
The assumption of negligible recession rate on the fluid system is used throughout this work.\\
\\
Equation \eqref{Conduction} for the solid is transformed similarly:
\begin{equation}
    \frac{\partial T_s}{\partial t} - v_s \frac{\partial T_s}{\partial x_2} = \alpha_s \nabla^2 T_s
\label{Conduction moving}
\end{equation}
No simplifying assumption is made for the solid to remove the surface velocity.

\section{Linearization} \label{chapter:linearization}
The governing equations, in the uniformly receding coordinate frame, $\bar{x}_i$, are linearized about a time-invariant parallel baseflow.
In general, a given quantity $\vphi(\bar{x}_i,t)$ is represented as 
\begin{equation}
    \vphi(\bar{x}_i,t) = \overline{\vphi}(\bar{x}_i) + \vphi^\prime(\bar{x}_i,t),
\end{equation}
where $\overline\vphi$ denotes the baseflow and $\vphi^\prime(\bar{x}_i,t)$ is a fluctuation.
Here, we re-label the Cartesian coordinate system $\bar{x}_i = (\bar{x}_1, \bar{x}_2, \bar{x}_3)$ to $\bar{x}_i = (x, y, z)$ for notational convenience.
For the local stability analysis considered herein, the baseflow depends only on $y$, while the fluctuations take the form
\begin{equation}
    \vphi^\prime(x_i,t) = \hat\vphi(y) e^{i\left(\alpha x + \beta z - \omega t\right)},
\end{equation}
where $\alpha$ and $\beta$ are the wavenumbers in the streamwise and spanwise directions, respectively, and $\omega$ is the frequency.
The orientation of the perturbation is illustrated graphically in the schematic provided in Figure \ref{fig:linear stability diagram}.

\subsection{Fluid}
The governing fluid equations in \eqref{Navier-Stokes} can be written compactly as $\partial_t\vQ = \vf(\vQ)$.
By expanding this equation about a steady-state solution $\overline{\vQ}$ and neglecting higher-order terms, a linear time-invariant system is obtained
\begin{equation}
     \frac{\partial \vQ'}{\partial t} = \underbrace{D\vf(\overline{\vQ})}_{ \coloneqq A(\overline{\vQ})}\vQ^\prime,
    \label{Linear Fluid}
\end{equation}
where $\vQ^\prime$ denotes a fluctuation around $\overline{\vQ}$.
Applying the modal ansatz to the linearized system~\eqref{Linear Fluid} gives
\begin{equation}
    -i\omega \hat{q} = \widetilde A\left(\overline{\vQ},\alpha,\beta \right) \hat{q},
\end{equation}
where the linear operator $\widetilde{A}$ is defined implicitly by means of the Fr\'echet derivative and the Fourier transform in Appendix \ref{appendix B}.

\subsubsection{Baseflow}
The baseflow is a self-similar compressible boundary layer with a non-reacting binary gas mixture.
The equations are provided by \citet{lysenko2021influence} and are repeated here for completeness.
In particular,
\begin{equation}
   \begin{array}{c}
    \frac{d}{dy}\left(\mu \frac{dU}{dy} \right) + F \frac{dU}{dy} = 0 , \quad 
    \frac{dq}{dy} = F \frac{dh}{dy} + \left(\gamma_e-1\right) M_e^2 \mu \left(\frac{dU}{dy} \right)^2 , \quad
    \frac{d j_1}{dy} = F\frac{d c_1}{dy}
    \end{array}
  \label{Self-Similar}
\end{equation}
where $2 \frac{dF}{dy} = \rho U$, $q=\kappa \frac{dT}{dy} + j_1\left(h_1 - h_2\right)$ and $j_1 = -\rho D_{12} \frac{dc_1}{dy}$. 
The equations have been non-dimensionalized by the boundary layer edge quantities, with the Blasius length scale as the reference length $\delta_{Bl} = \sqrt{x \mu_e/U_e \rho_e}$.
The slow-recession assumption discussed in Section \ref{subsection:receding surface} is present in these equations.
The boundary conditions are found from the no-slip wall and the surface balances from Equations \eqref{mass balance} and \eqref{energy balance}. With $c_1$ representing the sublimating species mass fraction, the surface boundary conditions in non-dimensional form are:
\begin{equation}
   \begin{array}{c}
    U(0) = 0 , \quad
    F(0) = -f_w = \frac{(\rho v)_f}{\rho_e u_e} \Rey \\
    f_w h_{sg} + \left(-\kappa \frac{dT}{dy}\right)_f = \left(- \kappa \frac{dT}{dy} \right)_{s} , \quad
    f_w \left(1-c_{1,w} \right) = -\rho_w D_{12,w} \frac{d c_1}{dy}\rvert_w,
    \end{array}
  \label{Self-Similar wall}
\end{equation}
where $\Rey = \rho_e u_e \delta_{Bl}/\mu_e$. The far-field conditions are $U(\infty) \to 1$, $T(\infty) \to 1$, and $c_1(\infty) \to 0$.
The equations are solved using a shooting method with iterations on the undetermined wall boundary conditions.

\subsubsection{Boundary Conditions}
Since the governing equations are solved on the nominally receding surface $\overline{v}_s$, the effects of perturbations to the surface resulting from $v_s^\prime$ must be accounted for. 
Taking the surface position to be at $y=\eta^\prime$, where the nominal surface is at $y=0$, the full interface boundary conditions resulting from no-slip, zero pressure gradient, and the interface mass/energy balances are:
\begin{equation}
\label{eq:nonlinear wall bcs}
    \begin{aligned}
        u(\eta^\prime) &= w(\eta^\prime) = 0,\quad v(\eta^\prime) = -\dot{m} / \rho_s \\
        \frac{\partial p}{\partial y}(\eta') = 0,\quad \frac{\partial T}{\partial y}(\eta') &= \frac{1}{\kappa_f}\left(\dot{m} H_{sg} +  \kappa_s \frac{\partial T_s}{\partial y} (\eta') \right),\enspace \text{and} \enspace \frac{\partial c_1}{\partial y} (\eta')=\frac{\dot{m}(1-c_1)}{-\rho D}.
    \end{aligned}
\end{equation}
Note that the fluid boundary conditions are coupled both to the interface and the solid state through the interface balances.
These conditions can be transformed from the unknown interface position, $y=\eta'$, to the nominal surface, $y=0$, through a first-order Taylor expansion:
\begin{equation}
    \phi(\eta^\prime) = \phi(0) + \frac{\partial \overline{\phi}}{\partial y} \bigg \rvert_0 \eta^\prime. 
    \label{eq:BC perturbation}
\end{equation}
\noindent Utilizing this transformation, the boundary conditions are then numerically enforced at the nominal surface, $y=0$, as:
\begin{equation}
    \begin{aligned}
        u(0) = &- \frac{\partial \bar{u}}{\partial y} \big |_0\eta', \quad v(0) = -\dot{m}^\prime / \rho_s, \quad w(0) = 0,\\
        p(0) = p(\eta'), \enspace \frac{\partial T}{\partial y}(0) &= \frac{\partial T}{\partial y}(\eta') - \frac{\partial^2 \bar{T}}{\partial y^2} \bigg |_0 \eta', \enspace \text{and} \enspace \frac{\partial c_n}{\partial y}(0) = \frac{\partial c_n}{\partial y}(\eta') - \frac{\partial^2 \bar{c}_n}{\partial y^2} \bigg |_0 \eta'.
    \end{aligned}
\end{equation}
This method of applying boundary conditions to the nominal surface position by linear expansion has been used on similar problems \citep{lees1972stability, lekoudis1976compressible}.
The far-field is treated with a sponge region and zero-gradient boundary conditions to eliminate reflections \citep{bodony2006analysis}.
To verify the far-field treatment, the domain geometry and sponge strength were varied and the eigenvalues and eigenvectors analyzed in Appendix \ref{appendix C}.
\subsection{Solid}
\label{sec:solid}
The linearization and modal ansatz are applied to the solid equation \eqref{Conduction moving}, giving:
\begin{equation}
\frac{d^2 \hat{T}_s}{d y^2} + \frac{\overline{v}_s}{\alpha_s}\frac{d \hat{T}_s}{d y} + \left(\frac{i \omega}{\alpha_s} - \chi^2 \right)\hat{T}_s = 0
\end{equation}
where $\chi^2=\alpha^2 + \beta^2$.
This is a second-order ordinary differential equation with constant coefficients. 
The analytical solution is:
\begin{equation}
\hat{T}_s = c_1 e^{m_1 y} + c_2 e^{m_2 y}, \quad m_{1,2} = \frac{1}{2}\left[ -\frac{\overline{v}_s}{\alpha_s} \pm \sqrt{\left( \frac{\overline{v}_s}{\alpha_s} \right)^2 - 4\left(\frac{i \omega}{\alpha_s} - \chi^2 \right)} \right] := q \pm r
\label{eq: m12}
\end{equation}
Considering a solid of thickness $h$, an isothermal far-field boundary condition at $y=-h$ gives:
\begin{equation}
\frac{d \hat{T}_s}{d y}(\hat{\eta}) = \left(q + \frac{r}{\tanh\left(rh\right)} \right) \hat{T}_s(\hat{\eta}).
\label{eq:full conduction solution}
\end{equation}
The total solid temperature gradient on the deformed surface is obtained using the expansion in \eqref{eq:BC perturbation}:
\begin{equation}
\frac{d T_s}{dy}(\eta') = \left( \frac{d \bar{T}_s}{dy}\bigg|_0 + \frac{d^2 \bar{T}_s}{dy^2} \bigg|_0 \eta' \right) + \left(q + \frac{r}{\tanh\left(rh\right)} \right) \left[ T_s(\eta') - \bar{T}_s(0) -  \frac{d \bar{T}_s}{dy} \bigg|_0 \eta' \right]
\label{eq:full conduction}
\end{equation}

\subsubsection{Baseflow}
Evaluation of \eqref{eq:full conduction} requires the first and second derivatives of $\overline{T_s}$ at the surface. For a baseflow with an isothermal wall condition, the surface temperature gradients in the solid are readily found from the energy balance in \eqref{energy balance - simple} and steady solution from \eqref{Conduction moving}.

\subsubsection{Boundary Conditions}
The analytical boundary condition for $d \hat{T}_s / dy$ given by \eqref{eq:full conduction solution} would lead to a nonlinear eigenvalue problem, where the eigenvalues are nonlinear functions of $\alpha$, $\beta$ and $\omega$.
To simplify, a quasi-steady assumption is made where the solid response occurs much faster than the surface response, and the solid is approximately at a steady-state.
Consequently, both $\overline{v}_s/\alpha_s$ and $\omega/\alpha_s$ are assumed much smaller than $\chi$, and taking $h \to \infty$, \eqref{eq:full conduction solution} simplifies to
\begin{equation}
\frac{d \hat{T}_s}{d y}(\hat{\eta}) = \chi \hat{T}_s(\hat{\eta}).
\end{equation}
This boundary condition is now linear for a fixed oblique angle, $\psi$, as $\chi = \alpha\sqrt{1+\tan^2\left(\psi\right)}$.

\section{Results} \label{chapter:Results}
Self-similar baseflows of a binary mixture boundary layer over a sublimating surface are computed for a range of freestream and wall temperature conditions. 
The freestream condition ranges are $M=[2,3,4]$, $Re=[500, 785, 5000]$, $\tilde{T}_0=T_0/T_{TP}=[0.65,0.88,1.10]$, and $\tilde{p}_0=p_0/p_{TP}=[0.2,2,20]$.
The freestream gas consists of air with $\gamma=1.4$ and $R=287~\mathrm{J/kg \cdot K}$.
The wall temperature is chosen relative to the adiabatic temperature, $T_{ad}$, defined as the temperature at which the solid interface is adiabatic, $d\overline{T}_s/dy=0$.
The wall temperature ratio is then defined as $T_r = T_w/T_{ad}$.
The sublimating material is camphor ($C_{10}H_{16}O$) with molar mass $MW=152.23~\mathrm{g/mol}$.
The relevant properties of camphor were presented in \cite{zibitsker2024validation} and are repeated here for completeness.
The solid properties are $\rho_s=990~\mathrm{kg/m^3}$, $\kappa_s=0.2~\mathrm{W/m \cdot K}$, and $c_{p,s}=271.2~\mathrm{J/mol \cdot K}$.
The vapor and transport properties are $T_{TP}=453.1~\mathrm{K}$, $p_{TP}=0.514~\mathrm{bar}$, $\sigma=6.87~\mathrm{\mathring{A}}$, $\varepsilon/k_B=562~\mathrm{K}$, and $c_{p,g}=192~\mathrm{J/mol \cdot K}$.
The specific heat of the gas is treated as constant and was obtained from the NASA-7 polynomial in \cite{zibitsker2024validation} at $T=300~\mathrm{K}$.
The accommodation coefficient is $a_1=0.18$, and the enthalpy of sublimation is $h_{sg}=51.9~\mathrm{kJ/mol \cdot K}$.

\begin{figure}
    \centering
  \includegraphics[width=0.42\columnwidth]{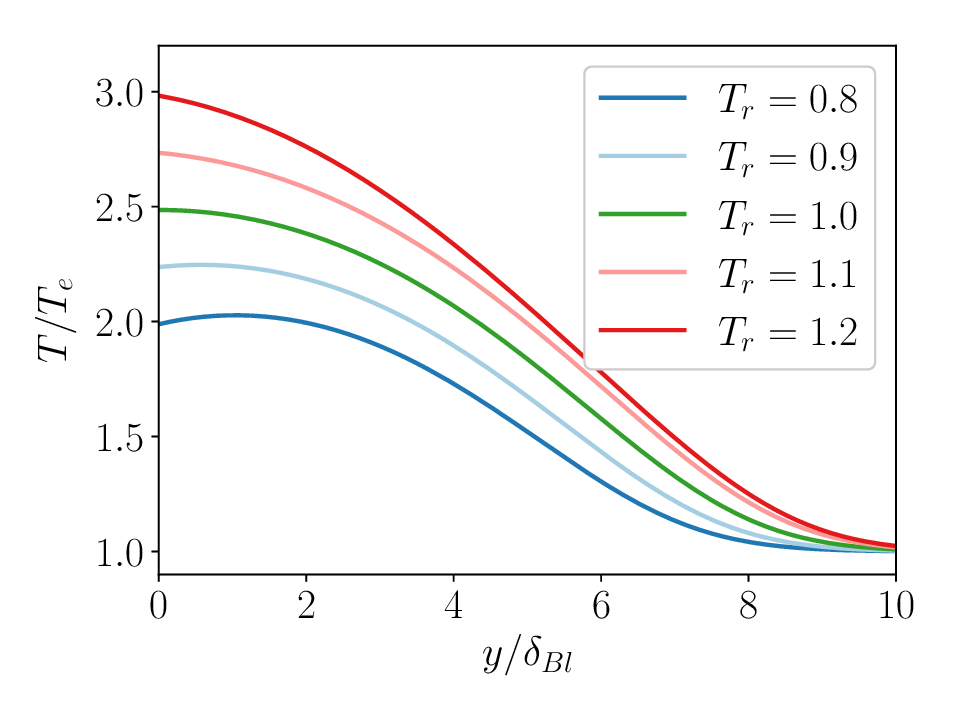}
  \includegraphics[width=0.42\columnwidth]{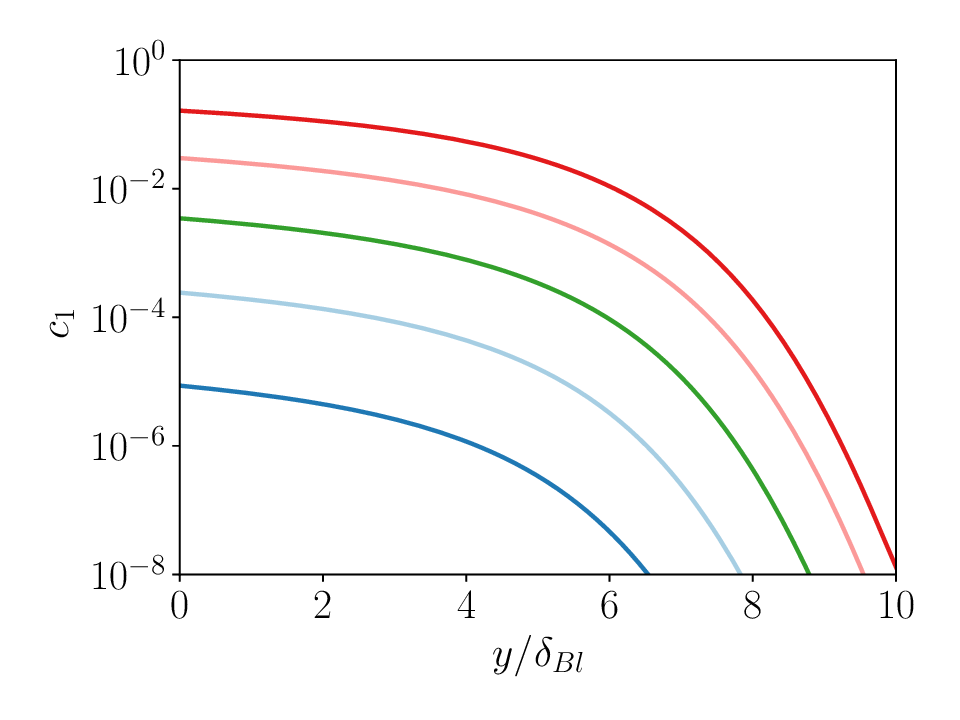}
  \caption{Self-similar baseflows with binary sublimation of camphor at $M=3$, $Re=785$, $\tilde{T}_0=0.65$, and $\tilde{p}_0=2$.}
\label{fig:baseflows-isothermal}
\end{figure}

Baseflow temperature and mass fraction profiles are shown in Figure \ref{fig:baseflows-isothermal} at $M=3$, $Re=785$, $\tilde{T}_0=0.65$, and $\tilde{p}_0=2$ for a range of wall temperatures ratios from $T_r=0.8-1.2$.
The profiles in Figure \ref{fig:baseflows-isothermal} show that for increasing wall temperature, the boundary layer thickness increases due to sublimation blowing. 
The amount of sublimation occurring for each case is illustrated by the vapor mass fraction $c_1$, with the $T_r=1.2$ case consisting of $16 \%$ sublimated vapor at the wall.
The case of $T_r=1.0$ corresponds to an adiabatic solid.

\subsection{Surface Mode Identification}
\begin{figure}
    \centering
    \includegraphics[width=0.42\textwidth]{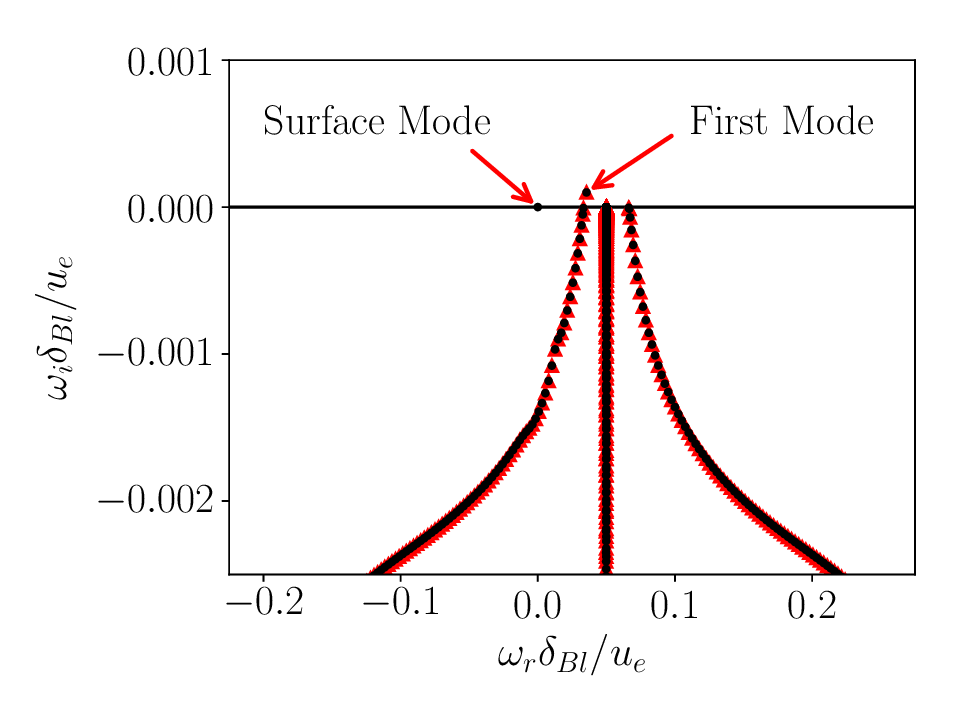}
    \includegraphics[width=0.42\textwidth]{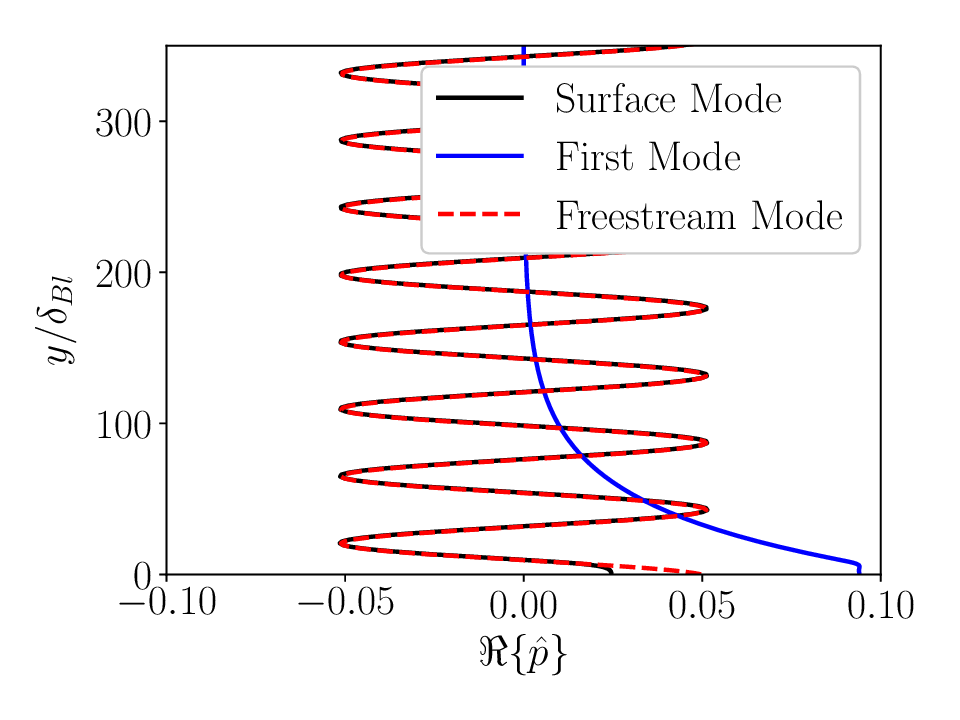} 
    \caption{Eigenvalues for a sublimating (black dots) and non-sublimating (red triangles) interface (left) and pressure component of eigenmodes (right). The baseflow conditions are $M=3$, $Re=5000$, $\tilde{T}_0=0.65$, $\tilde{p}_0=2$, and $T_r=1.0$.}
    \label{fig:surface_recession}
\end{figure}

The linearized governing equations with the quasi-steady conduction model are obtained for a given baseflow and the resulting temporal eigenvalue problem is solved.
The eigenvalues near the origin for the case of $M=3$, $Re=5000$, $\tilde{T}_0=0.65$, $\tilde{p}_0=2$, and $T_r=1.0$ are shown as black dots in Figure \ref{fig:surface_recession} for a wavenumber of $\alpha\delta_{Bl}=0.05$.
\\
\\
Observe from Figure \ref{fig:surface_recession} the presence of a first Mack mode instability, positioned between the slow acoustic and the entropy branches. 
This mode is also referred to as the Tolmien-Schlicting mode in incompressible flows.
Figure \ref{fig:surface_recession} also shows a single mode near the origin, labeled the surface recession mode, which appears only when recession effects are included in the linear system.
Under the conditions shown, this surface mode is stable ($\omega_i<0$)
To demonstrate the connection of this mode with surface recession, the eigenvalues were computed for the same problem without surface recession and are displayed in Figure \ref{fig:surface_recession} with red triangles, where the surface mode is now absent.

To further illustrate the characteristics of the surface mode, the Jacobian is transformed to primitive variables and the real part of the pressure component of the eigenvectors are shown in the right panel of Figure \ref{fig:surface_recession}.
An overlay of the phase-aligned eigenvector for the surface mode using the uniform freestream linear system is included for comparison, along with the first Mack mode.
The freestream system is provided in Equation 3.15 of \citet{ozgen2008linear}.
The surface mode exhibits oscillations in the freestream that are consistent with the freestream eigenmode and can be interpreted as Mach waves emanating from a wavy surface.
The wavenumber for Mach waves is computed geometrically as
\begin{equation}
    \alpha_y = \chi \cot\left(\sin^{-1}(1/M_n) \right),
\end{equation}
where $M_n=M\cos(\psi)$ is the Mach number normal to the perturbed surface.
For the case analyzed in Figure \ref{fig:surface_recession}, the Mach wave analysis predicts a wavenumber of $\alpha_y \delta_{Bl}=0.14142$ and the freestream eigenvalues predict a wavenumber of $\alpha_y \delta_{Bl}=0.14142$ with a decay rate of $1.37 \times 10^{-5}$.

Finally, to justify the use of the quasi-steady conduction model in this analysis, an a posteriori comparison is made between the terms in Equation \eqref{eq: m12}. 
The surface mode in Figure \ref{fig:surface_recession} is located at $\omega=(6.85-72.20i) \times 10^{-7}$ rad/s.
Evaluating the terms in Equation \eqref{eq: m12} gives $m_{1,2}/\alpha = 0.0004 \pm \left( 1.005 - 0.0005 i \right)$, validating the quasi-steady assumption that $m_{1,2}\approx \alpha$.

\subsection{2D Surface Modes}
\begin{figure}
\centering
  \includegraphics[width=0.9\columnwidth]{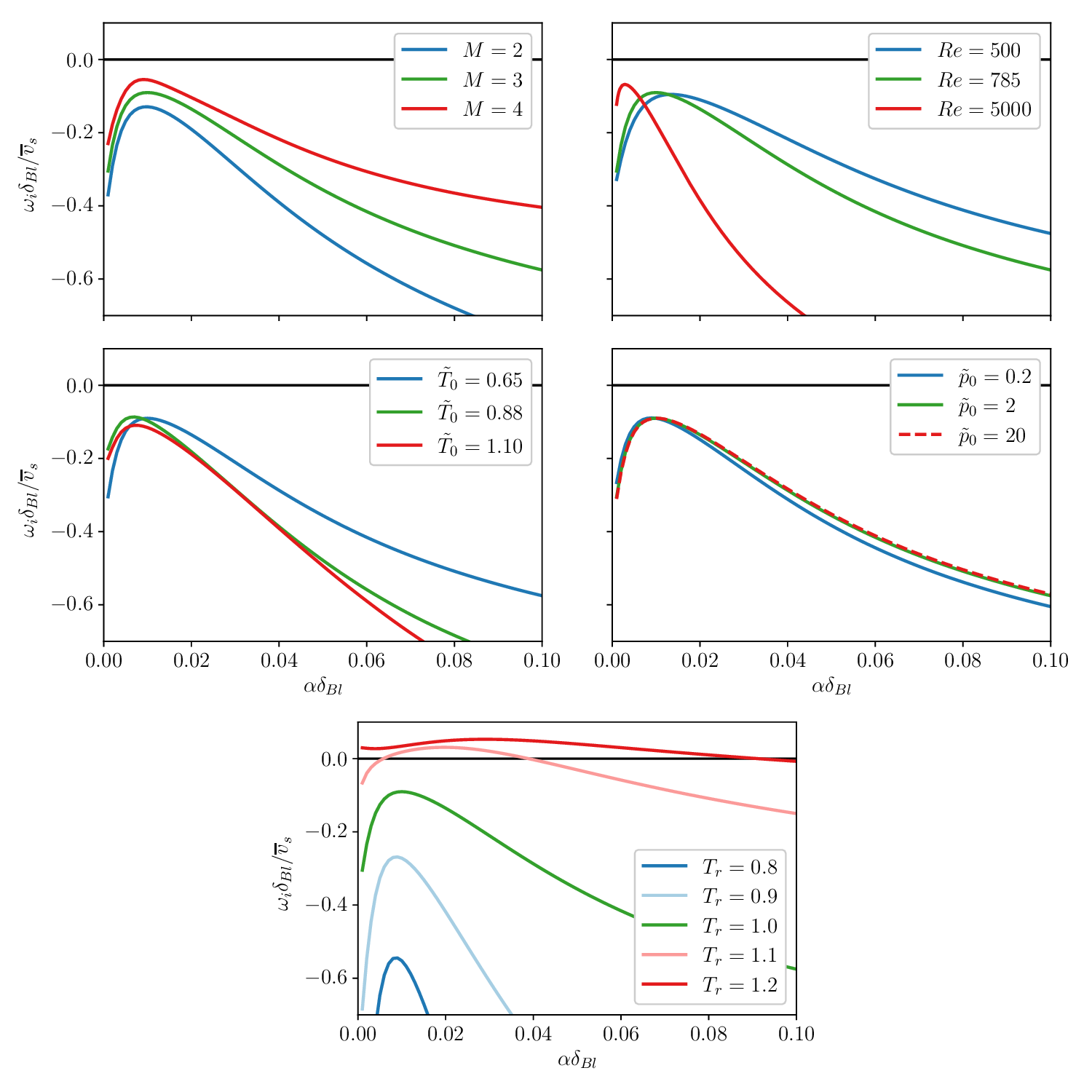}
  \caption{Growth rates for 2D surface modes for variations in the baseflow conditions with respect to a common condition of $M=3$, $Re=785$, $\tilde{T}_0=0.65$, $\tilde{p}_0=2$, and $T_r=1.0$. The single condition varied is indicated in the legend of each panel. The black lines mark neutral stability.}
\label{fig:2D growth}
\end{figure}

To study the effect of baseflow conditions on the surface mode, temporal growth rates for streamwise-2D perturbations ($\beta=0$) are computed for variations in the baseflows with respect to a common condition of $M=3$, $Re=785$, $\tilde{T}_0=0.65$, $\tilde{p}_0=2$, and $T_r=1.0$. The resulting growth rates as a function of wavenumber are shown in Figure \ref{fig:2D growth}.
The results in Figure \ref{fig:2D growth} show that variations in $M$, $Re$, $\tilde{T}_0$, and $\tilde{p}_0$ about the base condition do not destabilize the surface.
The effect of increasing $M$ trends towards instability, but a $M=5$ result cannot be obtained under the transport model due to the low freestream temperatures \citep{kim2014high}. 
The results in the bottom panel of Figure \ref{fig:2D growth} show that the surface is unstable for wall temperatures exceeding the adiabatic case ($T_r > 1.0$).
A critical wavenumber exists at which the temporal growth rate reaches a maximum for all cases, meaning a surface pattern with a specific wavelength may emerge on sublimating plates. 

\subsubsection{Neutral Stability}
\begin{figure}
    \centering
    \includegraphics[width=0.42\textwidth]{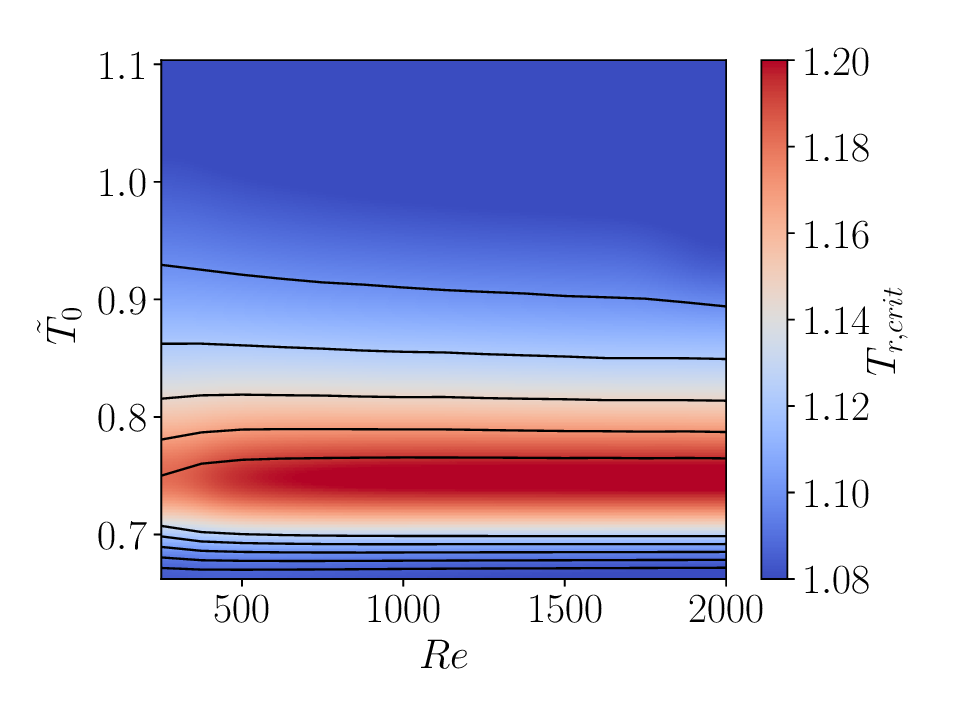} 
    \includegraphics[width=0.42\textwidth]{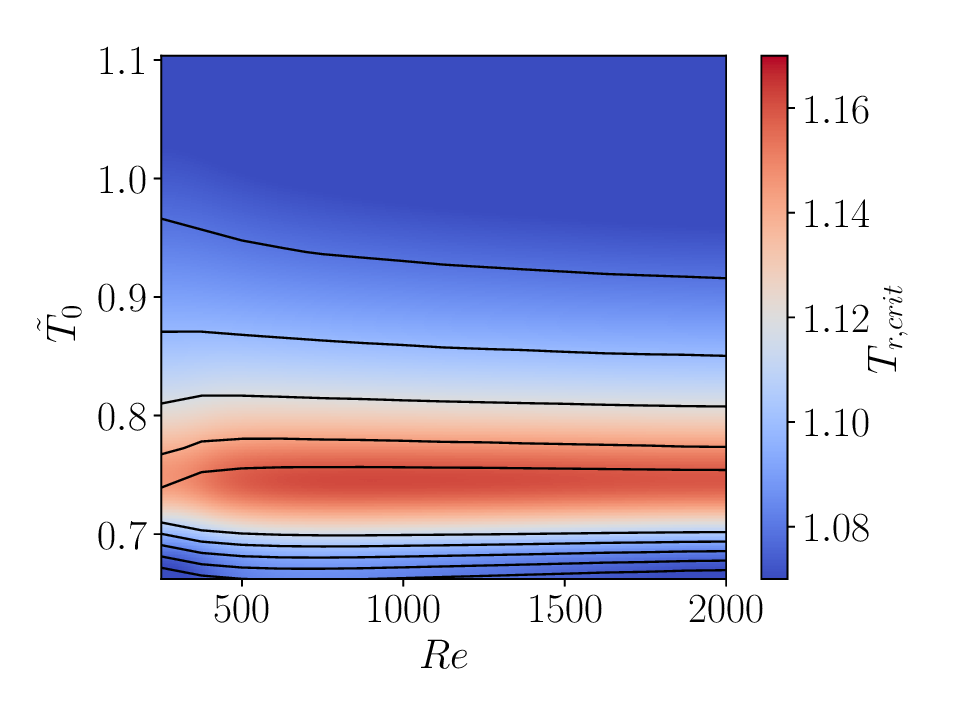} 
    \caption{Neutral stability maps of $T_{r,crit}$ for $M=3$ (left) and $M=4$ (right)}
    \label{fig:neutral_stability}
\end{figure}

To further explore the range of flow conditions that result in an unstable surface, the critical temperature ratio, $T_{r,crit}$, that gives a neutrally stable surface is computed for a range of $\tilde{T}_0$ and $Re$. The results for $M=3$ and $M=4$ are shown in Figure \ref{fig:neutral_stability}. Note that $\tilde{p}_0=2$ is held constant due to the insensitivity observed in Figure \ref{fig:2D growth}.
\\
\\
From Figure \ref{fig:neutral_stability} it is observed that, consistent with the trends in Figure \ref{fig:2D growth}, increasing $Re$ and $M$ have a slight destabilizing effect, resulting in decreased $T_{r, crit}$ needed to destabilize the surface.
The neutral stability maps also show an island of relative stability exists from $0.7 \leq \tilde{T}_0 \leq 0.8$, where the $T_{r,crit}$ needed to destabilize the surface increases.
The cause of this stabilizing effect is not known and further investigation of the mechanisms are needed.
Given that a $T_{r,crit}$ exists for a wide range of freestream conditions, and $T_{r,crit}>1$ for each case, a hot wall condition appears to be required to destabilize the surface.\\
\\
The results in Figures \ref{fig:2D growth} and \ref{fig:neutral_stability} confirm the existence of a surface instability due to differential ablation in flow over a sublimating surface, consistent with the findings of \cite{lees1972stability}.
However, this is the first time such an instability has been identified in a laminar boundary layer, with the necessary condition of a hot wall ($T_r>1.0$).
Furthermore, these laminar stability results are consistent with the experimental observations (summarized in Table \ref{tab:summary of experiments}) where laminar cold-walls ($T_r\leq 1.0$) are found to be stable.

\subsection{Oblique Surface Modes}
\begin{figure}
    \centering
    \includegraphics[width=0.43\textwidth]{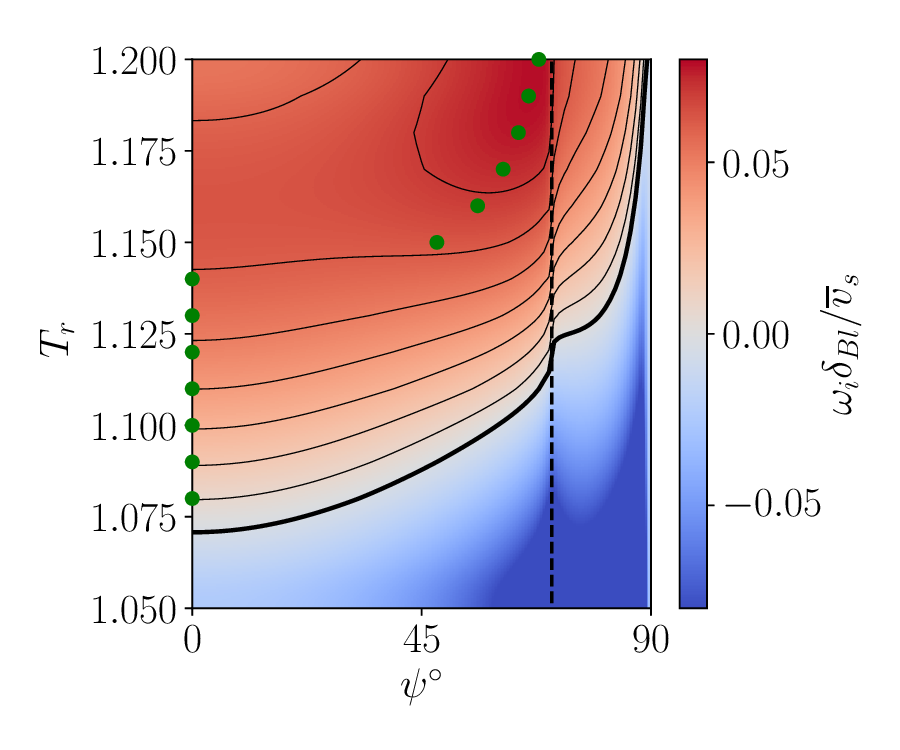} 
    \includegraphics[width=0.43\textwidth]{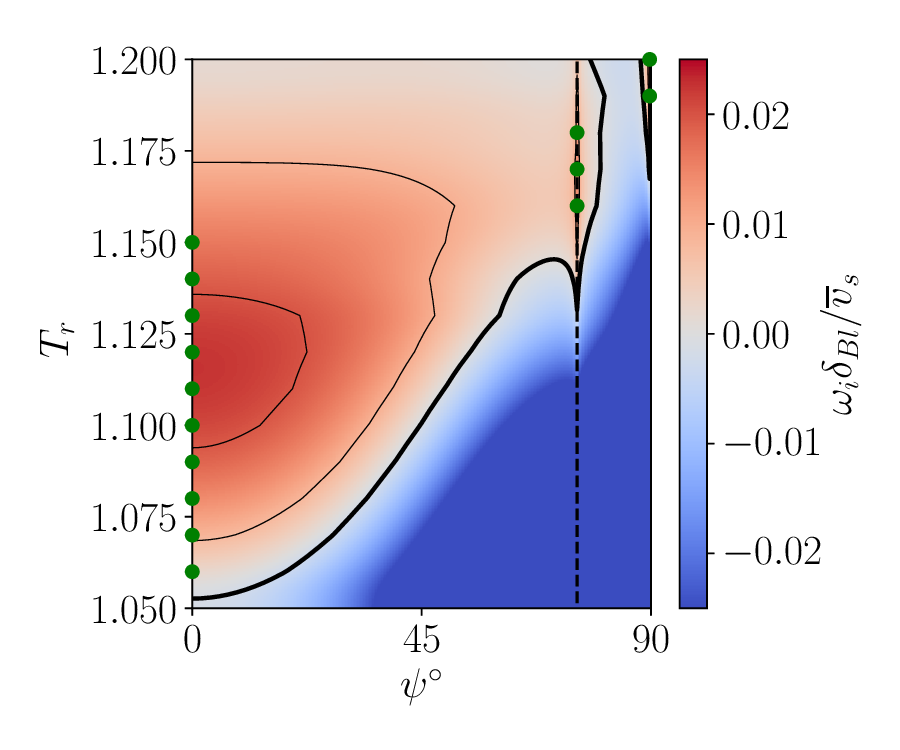} 
    \caption{Oblique surface growth rates at $\chi_{max}$ for $T_r=1.05-1.2$ at $M=3$ (left) and $M=4$ (right).
    The dashed lines mark the sonic angle, $\psi_s$.
    The thick solid black line marks neutral stability, while the thin black lines are iso-contours in 0.01 increments.
    The green dots mark the location of peak growth rate for each $T_r$, in 0.01 increments.
    }
    \label{fig:3D surface mode}
\end{figure}

Surface modes for oblique ($\beta \neq 0$) perturbations were computed across a range of oblique angles, defined as $\psi = \tan^{-1}(\beta/\alpha)$, at the streamwise wavenumbers corresponding to the maximum 2D growth rate, $\chi_{max}$. 
The temporal growth rates for both $M=3$ and $M=4$ conditions are presented in Figure \ref{fig:3D surface mode}.
The angle at which the normal Mach number reaches the sonic condition is indicated by the vertical dashed lines in Figure \ref{fig:3D surface mode}, and is given by $\psi_s = \cos^{-1}(1/M_e)$.
This angle is also the complement of the Mach angle. 
The results in Figure \ref{fig:3D surface mode} show that the streamwise-2D mode is the most unstable for both conditions until $T_r > 1.14$, beyond which an oblique mode near the sonic angle becomes more unstable.
Further increases in $T_r$ reveal a third unstable orientation of a spanwise mode near $\psi=90^{\circ}$ under the $M=4$ condition.
The rapid transition of the most unstable orientation from streamwise-2D to oblique to spanwise-2D suggests distinct instability mechanisms are present.
The importance of supersonic flow on surface instabilities is highlighted by the presence of the unstable sonic angle orientation, where as the oblique angle increases across the sonic angle, the flow normal to the surface perturbation transitions from supersonic to subsonic.

The presence of an instability at an oblique angle provides an intriguing connection to observed crosshatching patterns.
If linear differential ablation is the primary mechanism driving surface pattern formation, these results indicate that a single dominant wavelength and orientation will emerge, determined by the most unstable mode. 
Accordingly, sublimating materials exposed to supersonic, laminar, hot wall conditions are predicted to develop surface patterns at one of three preferred orientations: $\psi=0$, $\psi_s$, or $90^{\circ}$.
If crosshatching patterns are caused by linear instability mechanisms in turbulent flows, the sonic instability is a viable candidate as it contains a critical wavenumber and orientation, consistent with experimental observations.

\section{Conclusion}
The formulation for local linear stability analysis of ablating surfaces is presented and applied to a compressible boundary layer over a sublimating plate.
Temporal stability analysis is performed on sublimating baseflows, where hot walls are found to destabilize the laminar surface mode.
The unstable surface mode contains a critical wavenumber and orientation, aligning with experimental observations of crosshatching patterns in turbulent flows.
While the comparison between laminar and turbulent flows is not conclusive, it may suggest a common underlying mechanism and motivates continued study through linear stability methods.
The three-dimensional analysis reveals a transition in the dominant mode orientation from streamwise, to sonic, to spanwise as the temperature ratio, $T_r$ increases.
The presented linear framework can be used to further investigate surface instabilities under more conditions, including turbulent flows, and to explore additional mechanisms, such as inelastic deformation or liquid layer models. 
The analysis can also be extended to study non-modal effects and assess whether transient growth contributes to surface pattern formation.

\backsection[Funding]{This work was supported by the LDRD Program at Sandia National Laboratories. Sandia is managed and operated by NTESS under DOE NNSA contract DE-NA0003525.}

\backsection[Declaration of interests]{The authors report no conflict of interest.}

\bibliographystyle{jfm}
\bibliography{jfm}

\appendix
\section{Transport Model} \label{appendix A}
The single-species transport quantities are given by:
\begin{equation}
    \begin{array}{c}
    \mu_i = 2.6693 \times 10^{-6} \frac{\sqrt{T M_i}}{d_i^2 \Omega^{(2,2)}} , \quad
    \kappa_i = 0.0833 \frac{\sqrt{T/M_i}}{d_i^2 \Omega^{(2,2)}} , \\
    \quad \\
    D_{12} = 1.858 \times 10^{-7} \frac{\sqrt{T^3 \left(1/M_1 + 1/M_2 \right) }}{p d_{12}^2 \Omega^{(1,1)}}
    \end{array}
\label{Transport Model}
\end{equation}
where all terms are given in dimensional units as described in \cite{hirschfelder1964molecular}. 
The mixture transport coefficients are found using the Wilke mixing rule with Eucken correction for conductivity. 
For a binary mixture we have:
\begin{equation}
    \begin{array}{c}
    \mu = \frac{\mu_1}{1 + G_{12}(x_2/x_1)} + \frac{\mu_2}{1 + G_{21}(x_1/x_2)} , \quad
    \kappa = \frac{\tilde{\kappa}_1}{1 + 1.065\tilde{G}_{12}(x_2/x_1)} + \frac{\tilde{\kappa}_2}{1 + 1.065\tilde{G}_{21}(x_1/x_2)}
    \end{array}
\label{Wilke Mixing}
\end{equation}
where
\begin{equation}
    \begin{array}{c}
    G_{ij} = \frac{\left[1 + \sqrt{\mu_i/\mu_j} \left(m_j/m_i \right)^{1/4}\right]^2}{2^{3/2}\sqrt{1+ m_i/m_j}} , \quad
    \tilde{G}_{ij} = \frac{\left[1 + \sqrt{\kappa_i/\kappa_j} \left(m_j/m_i \right)^{1/4}\right]^2}{2^{3/2}\sqrt{1+ m_i/m_j}}
    \end{array}
\label{Wilke Mixing - details}
\end{equation}
and the Euken correction factor:
\begin{equation}
    \begin{array}{c}
    \tilde{\kappa}_i = Eu \kappa_i , \quad
    Eu = 0.115 + 0.354\frac{C_{pi}}{R_i}
    \end{array}
\label{Eucken}
\end{equation}
where $x_i=\frac{c_i/m_i}{c_1/m_1 + c_2/m_2}$ are the mass fractions.
Additional details are found in Chapter 10 of \cite{dorrance2017viscous}.
The collision integrals  for the kinetic theory are evaluated using the high-order curve fit provided by \citet{kim2014high}.

\section{Linearized System} \label{appendix B}
Letting $\Psi_{\alpha,\beta} := \mathcal{F}_{\alpha,\beta}(\psi)$ denote the $(\alpha,\beta)$-Fourier coefficient of some spatially-dependent function $\psi$, the Jacobian is numerically approximated from a nonlinear solver using the Fr\'echet derivative,
\begin{equation}
    \widetilde{A} \Psi_{\alpha, \beta} \approx \mathcal{F}_{\alpha,\beta} \left( \frac{\vf(\overline{\vQ} + \varepsilon \psi) - \vf(\overline{\vQ} - \varepsilon \psi)}{2 \varepsilon} \right)
\end{equation}
where $\vf$ is the right-hand-side of the nonlinear governing equations.

The Jacobian $\widetilde{A}$ obtained from the nonlinear solver is directly applicable to temporal eigenvalue analysis, but it requires recomputing the Jacobian for each new wavenumber.
Observe that the governing equations are at most second-order in space, with cross-derivatives. The Jacobian can be explicitly expressed in the form:
\begin{equation}
    -i\omega \hat{q} = \left( \alpha A_1 + \alpha^2 A_2 + A_3 + \beta A_4 + \beta^2 A_5 + \alpha \beta A_6 \right) \hat{q}
\label{Linear NS expanded}
\end{equation}
The individual complex matrices $A_i$ can be extracted by obtaining the full Jacobian, $\tilde{A}$, at six linearly independent combinations of wavenumbers.
Using this form, the Jacobian can be computed efficiently for all wavenumbers and be used for spatial stability analysis.

\begin{figure}
    \centering
    \includegraphics[width=0.5\textwidth]{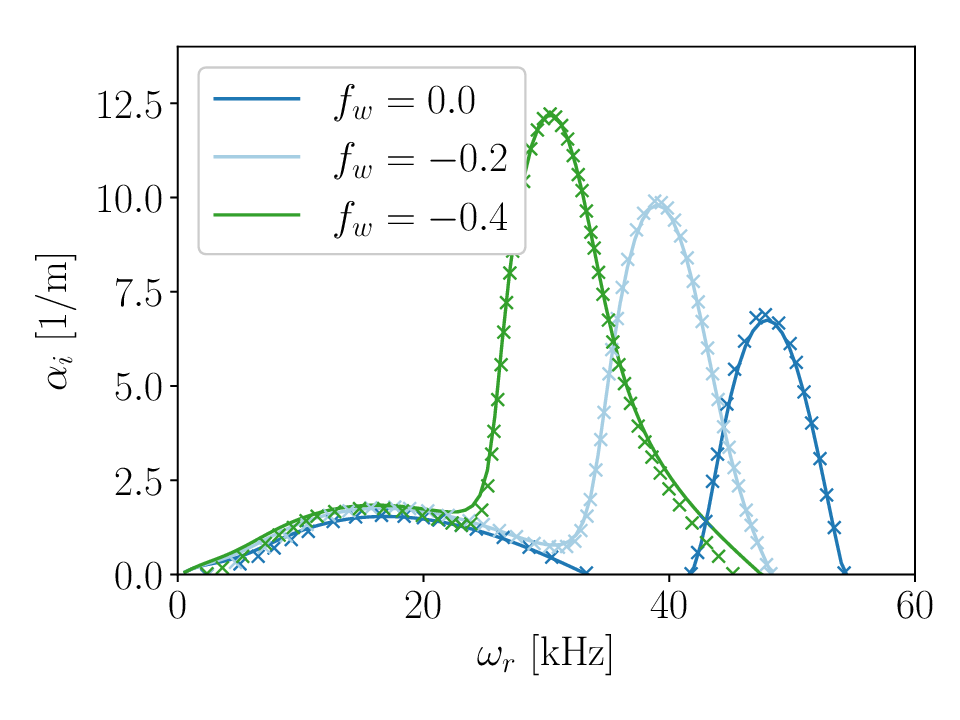}
    \caption{Spatial stability of compressible boundary layer with blowing. Symbols are results from \cite{ghaffari2010numerical}.}
    \label{fig:blowing verification}
\end{figure}

To verify this method, a spatial stability analysis of a $M=4.8$ boundary layer with blowing is performed and compared to the results of \citet{ghaffari2010numerical} in Figure \ref{fig:blowing verification}.

\section{Far-field verification} \label{appendix C}
\begin{figure}
    \centering
    \includegraphics[width=0.42\textwidth]{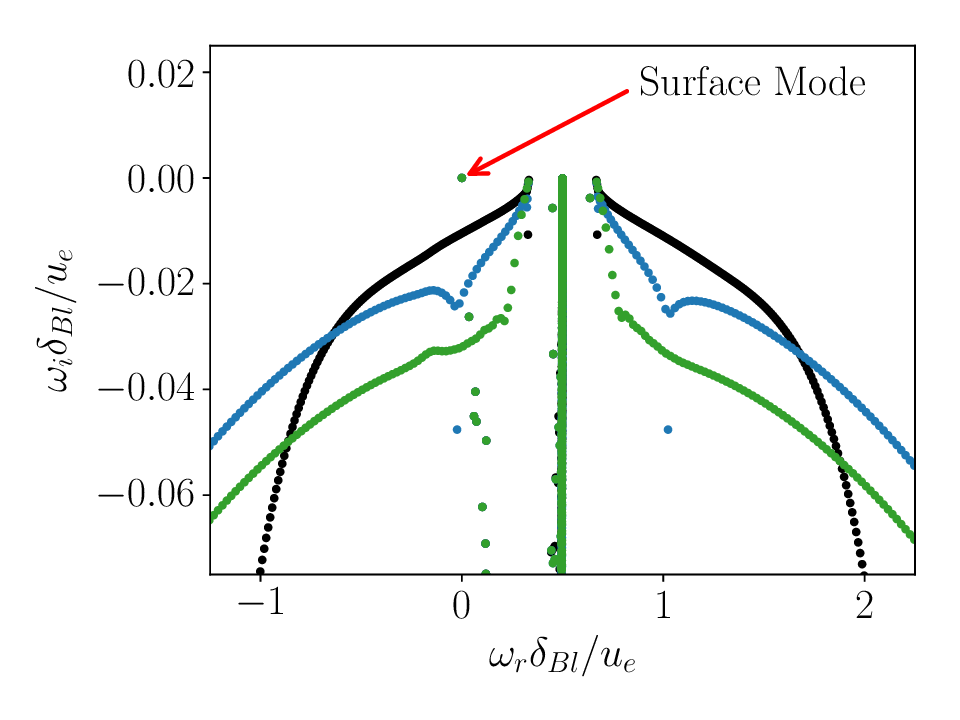} 
    \includegraphics[width=0.42\textwidth]{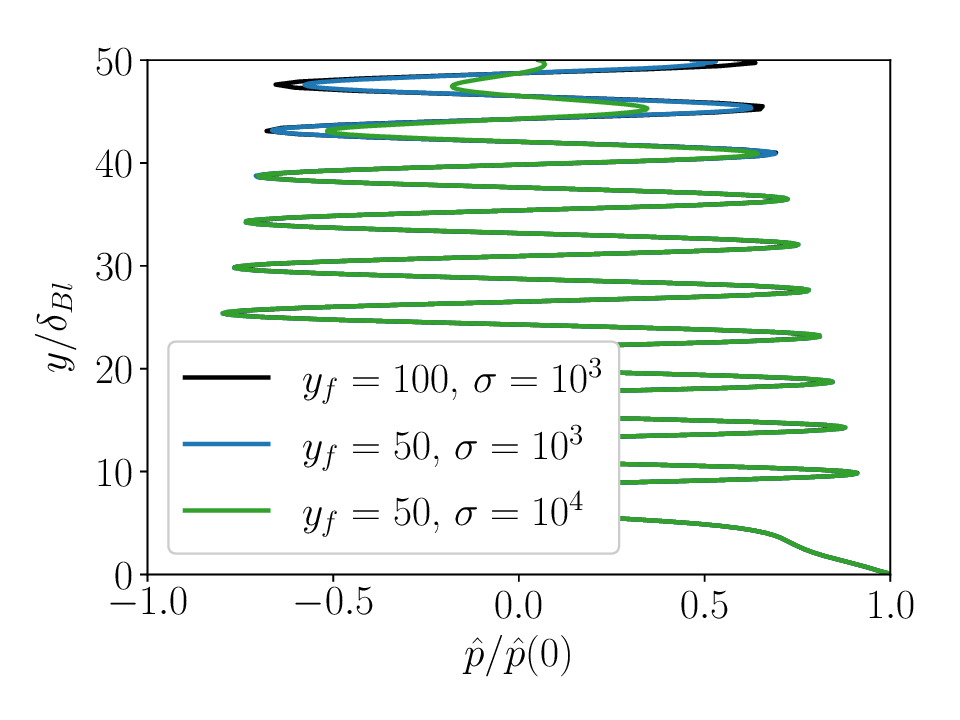} 
    \caption{Eigenvalues (left) and eigenmode pressure (right)}
    \label{fig:surface_recession_2}
\end{figure}

To verify the far-field boundary conditions are not influencing the surface stability results, a 2D temporal stability analysis was computed for varying far-field domain length $y_f$ and quadratic sponge strength $\sigma$. To illustrate the decay of the surface mode in the freestream, the conditions were chosen as $M=3.0$, $Re=785$, $\tilde{T}_0=0.65$, $\tilde{p}_0=2$, $T_r=1.0$, and $\alpha \delta=0.5$. The resulting eigenvalues and surface mode eigenvectors are shown in Figure \ref{fig:surface_recession_2}.
The results demonstrate that changing the domain and sponge properties have an influence on the eigenvalue spectra due to the changing resolution and numerical dissipation. However, a closer inspection of the surface mode finds a deviation of less than $0.1\%$ in the eigenvalue. Similarly, the eigenvectors show excellent agreement between the surface and the beginning of sponge region (beginning at $y=40$ when $y_f=50$). Once inside the sponge region, the amplitudes decay according to their sponge strengths, $\sigma$.
\end{document}